\documentclass[journal,letterpaper]{IEEEtran}

\usepackage{amsmath} 
\usepackage{amssymb}                         %输入实数集，复数集这样的符号
\usepackage[linesnumbered, ruled,norelsize]{algorithm2e}    %公式，见https://tex.stackexchange.com/questions/147598/how-to-use-the-algorithm2e-package-with-ieeetran-class
\usepackage{bm}                              %专门处理数学粗体的bm宏包
\usepackage{color}
\usepackage[pdftex]{graphicx}                %图片
\ifCLASSOPTIONcompsoc            %画subfigure
  \usepackage[caption=false,font=normalsize,labelfont=sf,textfont=sf]{subfig}
\else
  \usepackage[caption=false,font=footnotesize]{subfig}
\fi
\usepackage{setspace}%设置行距

\makeatletter
\newcommand{\removelatexerror}{\let\@latex@error\@gobble}
\makeatother

\begin{document}

\title{\huge Holographic Integrated Sensing and Communications: Principles, Technology, and Implementation}

\author{Haobo~Zhang,~\IEEEmembership{Student Member,~IEEE,}
        Hongliang~Zhang,~\IEEEmembership{Member,~IEEE,}
        Boya~Di,~\IEEEmembership{Member,~IEEE,}
        and~Lingyang~Song,~\IEEEmembership{Fellow,~IEEE}

\thanks{H. Zhang, H. Zhang, B. Di, and L. Song are with State Key Laboratory of Advanced Optical Communication Systems and Networks, School of Electronics, Peking University, Beijing 100871, China (e-mail: haobo.zhang@pku.edu.cn, hongliang.zhang92@gmail.com, \{diboya,lingyang.song\}@pku.edu.cn).}%
}

\maketitle

\begin{abstract}
Integrated sensing and communication~(ISAC) has attracted much attention as a promising approach to alleviate spectrum congestion. However, traditional ISAC systems rely on phased arrays to provide high spatial diversity, where enormous power-consuming components such as phase shifters are used, leading to the high power consumption of the system. In this article, we introduce holographic ISAC, a new paradigm to enable high spatial diversity with low power consumption by using reconfigurable holographic surfaces~(RHSs), which is an innovative type of planar antenna with densely deployed metamaterial elements. We first introduce the hardware structure and working principle of the RHS and then propose a novel holographic beamforming scheme for ISAC. Moreover, we build an RHS-enabled hardware prototype for ISAC and evaluate the system performance in the built prototype. Simulation and experimental results verify the feasibility of holographic ISAC and reveal the great potential of the RHS for reducing power consumption. Furthermore, future research directions and key challenges related to holographic ISAC are discussed. 
\end{abstract}


\IEEEpeerreviewmaketitle

\newcommand{\FigHeight}{2.4}

%%%%%%%%%%%%%%%%%%%%%%%%%%%%%%%%%%%%%%%%
\section{Introduction}
\label{s_introduction}
%%%%%%%%%%%%%%%%%%%%%%%%%%%%%%%%%%%%%%%%

Due to the explosion of wireless communication devices, spectrum congestion is becoming a severe problem, releasing increasing pressure on existing radar, communication, and other wireless systems. To cope with this issue, the concept of integrated sensing and communication~(ISAC) is proposed~\cite{liu2020jointradar}. It allows the sharing of hardware platform and spectrum between sensing and communication functions, which reduces the overall hardware cost and promotes spectrum efficiency.

In ISAC systems, massive multiple-input multiple-output~(MIMO) with a phased array is widely acknowledged as one of the vital techniques because the spatial diversity that it provides can be leveraged to improve the sensing accuracy and support high-speed communication. However, the following two bottlenecks have led to insufficient performance for phased array-enabled ISAC systems. \emph{First}, the power consumption of the phased array is relatively high due to extensive usage of complicated components such as phase shifters and power splitters. \emph{Second}, the gain of the phased array is limited as the antenna spacing in the phased array is typically half-wavelength and is hard to shrink due to the hardware implementation difficulties~\cite{wan2021terahertz}.

Recently, \emph{holographic radio} has been proposed as a new paradigm to address the above drawbacks of massive
MIMO enabled by phased arrays~\cite{pizzo2020spatially}. According to the vision of holographic radio, the antenna array in wireless systems is composed of a tremendous number of inexpensive antenna elements with low power consumption, tiny size, and ultra-close element spacing. Thus, by exploiting the high spatial diversity provided by numerous elements, high directive gain can be achieved for ISAC with an acceptable power consumption~\cite{deng2021reconfigurableholographic}. 

To fulfill this vision, reconfigurable holographic surfaces~(RHSs) are viewed as a promising solution~\cite{yurduseven2017design}. Specifically, the RHS is a type of metamaterial antenna whose sub-wavelength radiation elements are compactly arranged on a printed circuit board~(PCB). Due to the tunability of the radiation amplitudes of RHS elements, the radiation pattern of the RHS can be customized without the use of complicated phase shifters, thus significantly reducing the power consumption and cost~\cite{deng2021reconfigurable}. Besides, the feed of the RHS is embedded in the PCB, enabling the RHS to directly transmit or receive wireless signals with a low-profile structure. This is significantly different from reconfigurable intelligent surfaces~(RISs), another type of metamaterial antenna~\cite{zhang2021reconfigurable} which creates desired radiation pattern by reflecting the signals from the external feeds and tuning the reflection phase shifts of the elements\footnote{\textcolor{black}{The RIS and RHS are also different in terms of system model and hardware implementation. Specifically, since the feeds of the RIS are apart from the metasurface, there is an extra reflection path between the BS and the RIS in the RIS model compared with the RHS model. Another difference in the system model is that the phase shifts of RIS elements are tunable, while the radiation amplitudes are adjusted in the RHS. As for the implementation, the RIS and the BS antenna are deployed in different locations, while the RHS is compactly integrated with the BS and serves as the antenna of the BS.}}.

\begin{figure*}[!t]
  \setlength{\abovecaptionskip}{-0pt}
  \setlength{\belowcaptionskip}{-18pt}
  \centering
  \includegraphics[height=2in]{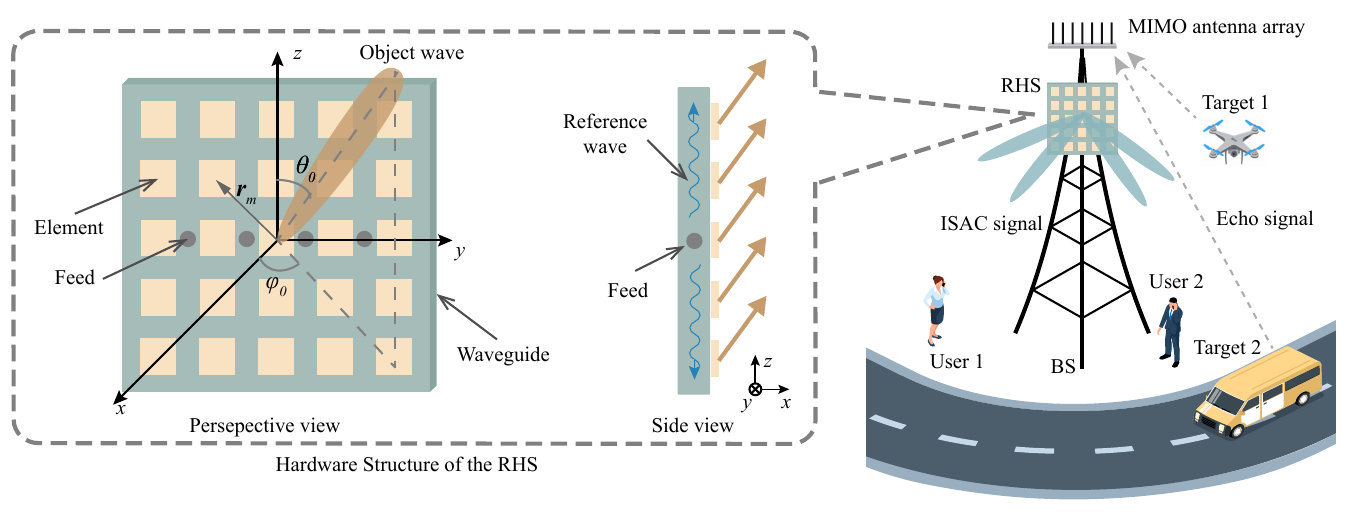}
  \caption{System scenario.}
  \label{f_scenario}
  \vspace{-4mm}
\end{figure*}

In this article, we propose \emph{holographic ISAC} where the holographic radio is used to further improve the performance of the ISAC systems. We design an RHS-enabled holographic beamforming scheme to realize holographic ISAC and evaluate its feasibility through experiment. Our contributions can be summarized below.

\begin{itemize}
  \item \textbf{Working Principles of RHSs:} The RHS is a planar antenna where the surface wave is first injected through feeds and then radiated to the free space through metamaterial elements. Since the electromagnetic responses of the metamaterial elements are tunable, the RHS does not rely on phase shifters to achieve the diversity of radiation patterns.
  \item \textbf{Beamforming Scheme for Holographic ISAC:} To serve communication users and detect targets at the same time, we propose a holographic beamforming scheme~\cite{zhang2022holographic}, where the communication streams and radar waveforms are first precoded at the base station~(BS) and then modulated by the RHS. The beamformers at the BS and the RHS are carefully selected to optimize the performance metrics for ISAC.
  \item \textbf{Implementation of Holographic ISAC:} Different from the theoretical investigation in~\cite{zhang2022holographic}, we build a hardware prototype consisting of an ISAC transceiver module, a communication user module, and a target module to validate the feasibility of holographic ISAC. Simulation and experiment results verify that with the aid of the RHS, the proposed platform is able to simultaneously sense the target and communicate with the user at a lower power consumption compared with the phased array-based platform. 
  \item \textbf{Challenges of Holographic ISAC:} In addition to the holographic beamforming schemes and prototype verification, we also discuss the relevant topics including the fundamental designs for the RHS, limitations and trade-offs of holographic ISAC, and the optimization of holographic ISAC transceiver.
\end{itemize}

The rest of this article is organized as follows. In Section~\ref{s_rhs}, the hardware structure and working principle of the RHS are described. A holographic beamforming scheme for ISAC systems is introduced in Section~\ref{s_hbi}. In Section~\ref{s_phis}, a hardware prototype of the holographic ISAC system is presented. The corresponding experimental results are reported in Section~\ref{s_pe}. In Section~\ref{s_frdkc}, future research directions and key challenges are discussed. Finally, conclusions are drawn in Section~\ref{s_c}.

\vspace{-1mm}
%%%%%%%%%%%%%%%%%%%%%%%%%%%%%%%%%%%%%%%%
\section{RHS: Hardware and Principles}
\label{s_rhs}
%%%%%%%%%%%%%%%%%%%%%%%%%%%%%%%%%%%%%%%%

\begin{figure*}[!t]
  \setlength{\abovecaptionskip}{-0pt}
  \setlength{\belowcaptionskip}{-18pt}
  \centering
  \includegraphics[height=1.8in]{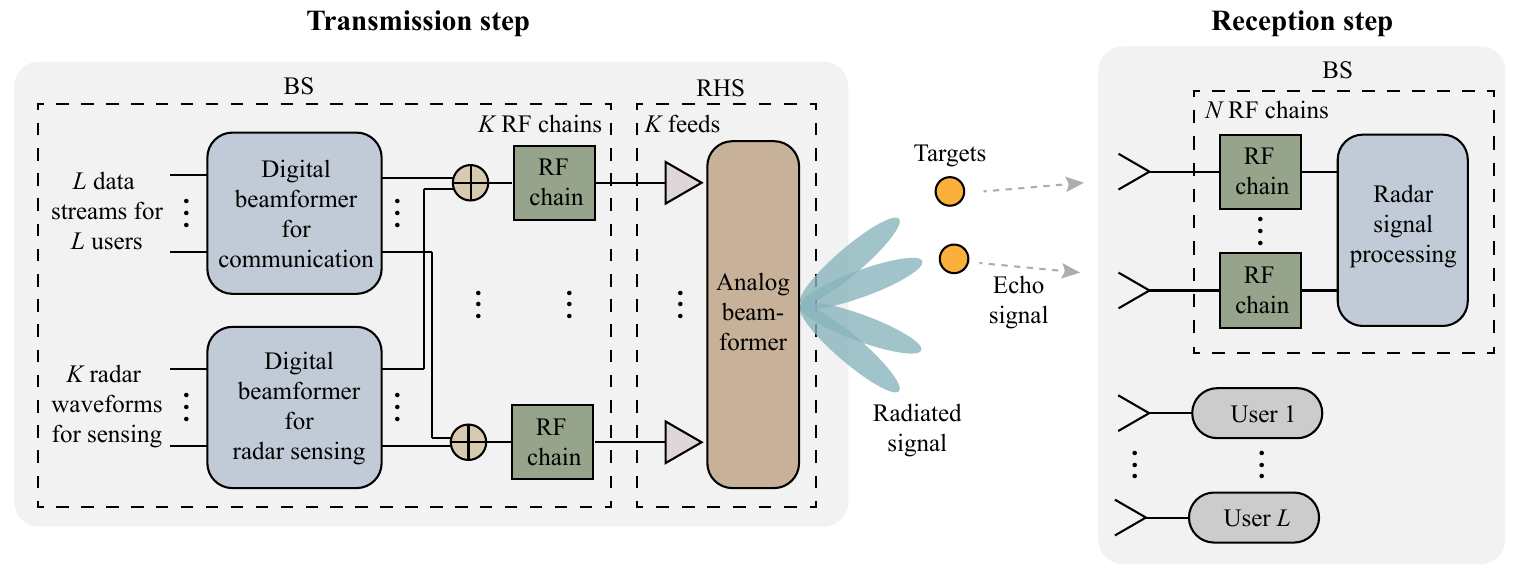}
  \caption{Block diagram of the proposed holographic beamforming scheme.}
  \label{f_scheme}
  \vspace{-4mm}
\end{figure*}

\vspace{-1mm}
\subsection{Hardware Structure}
\label{ss_hs}
\vspace{-1mm}

The RHS is a special type of leaky-wave antenna with controllable radiation patterns enabled by reconfigurable metamaterial elements. After the electromagnetic wave is injected into the antenna from feeds, it propagates in the waveguide and excites the metamaterial elements embedded on the waveguide to leak the energy into the free space. The radiation pattern of the RHS is formed by superposing the radiated signals, which are also referred to as the object waves, from all the elements. Thus, by controlling the radiation amplitudes of the elements, the radiation pattern can be altered to generate desired beams.

Fig.~\ref{f_scenario} illustrates the hardware structure of the RHS. It consists of three parts, i.e., the feeds, waveguide, and metamaterial elements, which are elaborated on as follows:

\begin{itemize}
  \item \textbf{Feed:} The feeds of the RHS are mounted at the bottom or the edge of the waveguide surface. The other side of each feed is connected to an RF chain, which will inject the electromagnetic waves, also called reference waves, into the RHS through the feed.
  \item \textbf{Waveguide:} The waveguide of the RHS is a planar medium where the reference wave propagates. The thickness of the waveguide is typically on the order of millimeters, leading to the ultra-thin characteristic of the RHS.
  \item \textbf{Metamaterial element:} The metamaterial elements arranged as a 1D or 2D array are laid on the top of the waveguide. Their electromagnetic responses can be independently adjusted by applying different bias voltages, enabling the RHS to control its radiation pattern.
\end{itemize}

Compared with the phased array, the structure of the RHS is much simpler because the RHS does not rely on complex feeding circuits and phase shifters. Moreover, the simpler structure also leads to lower power consumption of the RHS since power-consuming components such as amplifiers, phase shifters, and power splitters are not necessary for the RHS.

\vspace{-2mm}
\subsection{Working Principle}
\vspace{-1mm}
\label{ss_wp}

The working principle of the RHS is to first construct the interference pattern between the reference and object waves, and then excite the interference pattern recorded by the RHS to produce the desired radiation pattern~\cite{fong2010scalar}. The details of this principle are elaborated on as follows.

As shown in Fig.~\ref{f_scenario}, we consider an RHS with $M$ elements and $K$ feeds. Let $x_{\mathrm{ref}, m}$ and $x_{\mathrm{obj}, m}$ denote the reference and the object waves at the location of the $m$-th element, respectively. Here, the reference wave $x_{\mathrm{ref}, m}$ is the superposition of the reference waves from all the feeds, which is determined by the locations of the feeds, the location of the $m$-th element, and the propagation vector in the waveguide. The object wave $x_{\mathrm{obj}, m}$ whose main-lobe is pointing towards direction $(\theta_0, \phi_0)$ is determined by the location of the $m$-th element and the propagation vector towards direction $(\theta_0, \phi_0)$ in free space. 

Based on holographic interference principle~\cite{deng2021reconfigurable}, the interference pattern $x_{\mathrm{int}, m}$ at the location of the $m$-th radiation element can be expressed as $x_{\mathrm{ref}, m} x^*_{\mathrm{obj}, m}$. When the interference pattern is recorded by the RHS elements and excited by the reference wave $x_{\mathrm{ref}, m}$, the wave radiated by the RHS is in proportion to the desired object wave $x_{\mathrm{obj}, m}$, in this way the desired radiation pattern with main-lobe pointing towards direction $(\theta_0, \phi_0)$ is generated.

Since the interference pattern is complex, it cannot be directly recorded by the RHS element whose phase is not adjustable. Thus, we adopt a strategy of tunning the amplitudes of elements by using a real pattern, which is also referred to as a holographic pattern\footnote{\textcolor{black}{The information loss caused by using a real interference pattern rather than a complex interference pattern is very small. The readers may refer to~\cite{smith2017analysis} for more details.}}. The basic idea of this strategy is to radiate more energy when the reference wave and the object wave are in phase and radiate less energy when the waves are out of phase. Typically, the real part of the interference pattern $\text{Re}[x_{\mathrm{int}, m}]$ is chosen to construct this holographic pattern because $\text{Re}[x_{\mathrm{int}, m}]$ is negatively related to the phase difference between $x_{\mathrm{ref}, m}$ and $x_{\mathrm{obj}, m}$.

\vspace{-2mm}
%%%%%%%%%%%%%%%%%%%%%%%%%%%%%%%%%%%%%%%%
\section{Holographic Beamforming for ISAC}
\label{s_hbi}
%%%%%%%%%%%%%%%%%%%%%%%%%%%%%%%%%%%%%%%%

In this section, we first introduce the holographic beamforming scheme for a multi-user case, then discuss the performance metrics for holographic ISAC, and finally design the beamformers to optimize the ISAC performance.

\begin{figure*}[!t]
  \setlength{\abovecaptionskip}{-0pt}
  \setlength{\belowcaptionskip}{-18pt}
  \centering
  \includegraphics[height=2.5in]{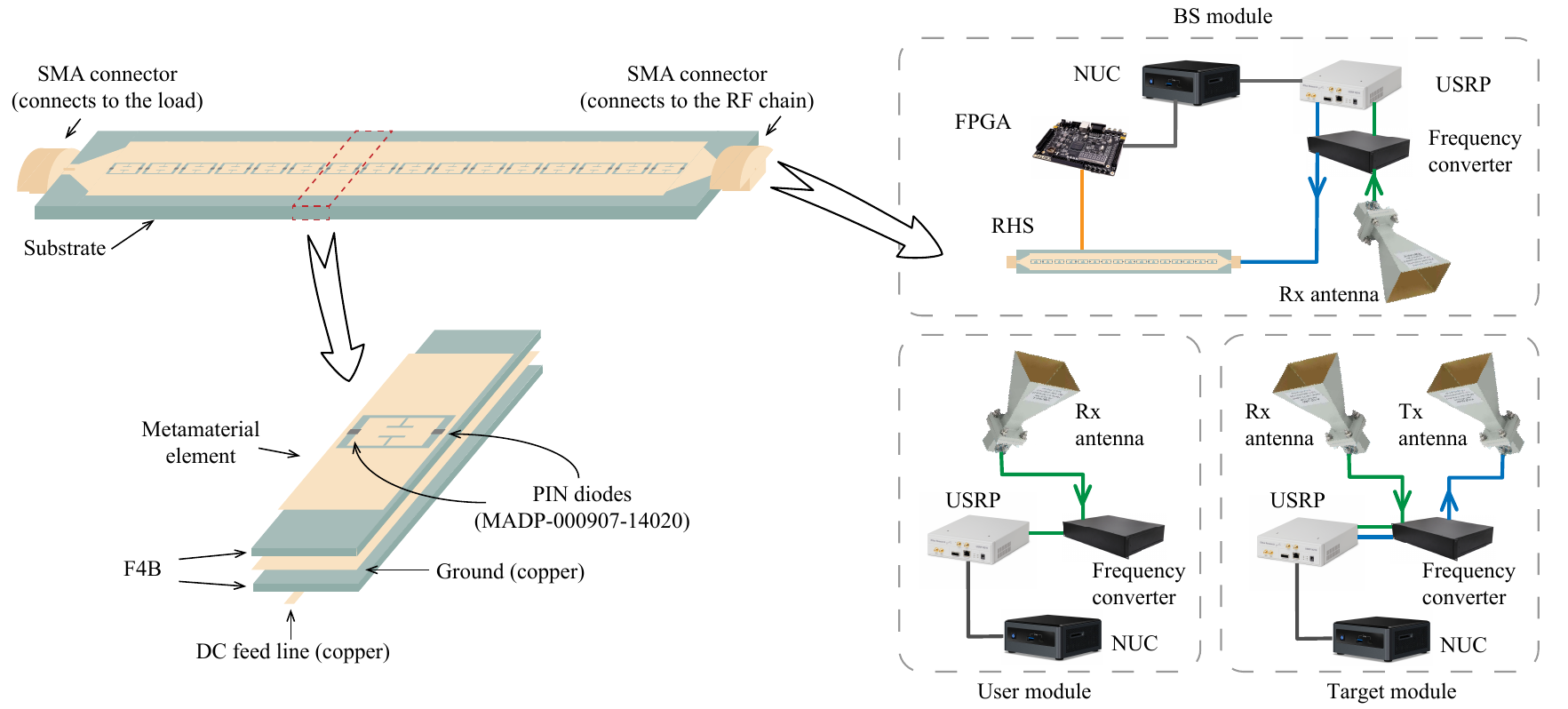}
  \caption{Illustration of the RHS-assisted ISAC prototype.}
  \label{f_prototype}
  \vspace{-4mm}
\end{figure*}

\vspace{-2mm}
\subsection{Holographic Beamforming Scheme}
\label{ss_hbs}

\textcolor{black}{As shown in Fig.~\ref{f_scenario}, the proposed ISAC system consists of a BS, an RHS\footnote{\textcolor{black}{The number of RHS feed can be $1$ in the system~\cite{zhang2018experimental}. The overall performances of the single-fed based system such as achievable data rates are lower than those of the multi-fed based system ($K > 1$) due to the lower flexibility on beam steering and shaping. However, the power consumption and cost of the system using a single-fed multi-beam antenna is lower than those using a multi-fed antenna because less hardware components such as RF chains are required in the system.}}, a MIMO antenna array, multiple downlink users, and multiple targets. The BS, RHS, and the MIMO antenna array are co-located with wired connections, unlike an usual RIS-assisted ISAC system where these terminals are not co-located. The BS acts as a terminal that feeds signals to the RHS for transmission and processes the echoes received via the MIMO antenna array. The signal transmission is only via the RHS, and the MIMO antenna array is used solely for the purpose of receiving the echo signals from the targets. In order to send different data streams to the downlink users and sense multiple targets at the same time, we propose the holographic beamforming scheme for the RHS-aided ISAC system, where the BS and the RHS perform digital and analog beamforming, respectively, to transmit ISAC signals.}

\textcolor{black}{Specifically, we divide the timeline of the scheme into cycles, and each cycle contains two steps, i.e., the transmission and reception steps. In the transmission step, the BS and the RHS cooperate with each other to conduct transmit beamforming and emit ISAC signals. In the reception step, the ISAC signals are recorded by the communication users for decoding, and the echo signals reflected by the targets are received by the MIMO antenna array for target sensing. The block diagram which shows the signal chain in each cycle is illustrated in Fig.~\ref{f_scheme}, which is further described in the following.}

\begin{itemize}
  \item \textcolor{black}{\textbf{Transmission step:} In this step, the ISAC beams that simultaneously serve communication users and detect targets are generated and transmitted. Specifically, $L$ data streams and $K$ radar waveforms are first separately precoded by the BS via different digital beamformers to generate $K$ precoded data streams and $K$ precoded radar waveforms. The $k$-th precoded data stream and radar waveform are added together, and $K$ added signals can be produced in total. These signals are then sent to the $K$ RF chains, and the RF chains use the input baseband signals to modulate the carrier signals and deliver the modulated signals to the feeds of the RHS. The injected signals from the feeds are converted to the radiation signals through the waveguide and metamaterial elements, where the signals are processed by the analog beamformer that determines the radiation amplitudes of the RHS elements.}
  \item \textcolor{black}{\textbf{Reception step:} The operations of communication and sensing are performed parallel in this step. To be specific, the communication users receive the ISAC signals transmitted by the RHS to retrieve their data streams, while the BS receives the echo signals reflected by the targets via the MIMO antenna array and then performs radar signal processing.}
\end{itemize}

\textcolor{black}{Note that the RHS needs to transmit multiple beams in different directions in order to simultaneously serve users and sense targets. Specifically, some of the emitted beams are directed at users to convey communication information, and others are utilized to sense targets. To generate these beams, we first create multiple holographic patterns. For each desired direction, $K$ patterns can be generated, each corresponding to one feed. Thus, the number of the generated patterns is equal to the product of the number of the directions and the number of the feeds. Next, all the holographic patterns are weighted superposed to derive the analog beamformer~\cite{deng2022HDMA}. The weights of the holographic patterns have to be optimized to promote the ISAC performance, which is discussed in the following subsection. It should be emphasized that the number of beams can be different from the number of feeds. For example, we can generate multiple beams with only one feed by using the above technique.}

\vspace{-4mm}
\subsection{ISAC Performance Metrics}
\label{ss_ipm}
\vspace{-1mm}

Since the ISAC tasks include sensing and communication, the related performance metrics can be broadly categorized into two types, i.e., sensing and communication metrics.

\emph{Sensing performance metrics:} Typically, the concept “sensing” has connotations of detection, estimation, and recognition. Specifically, detection refers to the process of deciding whether a target exists or not. Estimation means the judgment about the values of target parameters such as distance and velocity. And recognition refers to the act of identifying what the sensed target is. As the meanings of detection, estimation, and recognition vary, their performance metrics are different. For example, the performance of detection can be evaluated by detection probability (the probability of making a correct decision on the existence of the target), while the estimation performance is measured by mean squared error (MSE, the average value of the squared error between the true and the estimated values of a target parameter).

\emph{Communication performance metrics:} Similar to sensing tasks, the communication tasks also have different metrics. Two widely used metrics are channel capacity and bit error ratio~(BER). Channel capacity is defined as the maximal mutual information of the channel, and the capacity of an additive white Gaussian noise~(AWGN) channel can be calculated by the well-known Shannon formula. In contrast, BER means the percentage of the error bits in all the received bits, which measures the reliability of the communication systems.

\vspace{-4mm}
\subsection{Holographic Beamformer Optimization}
\label{ss_hbo}
\vspace{-1mm}

\textcolor{black}{In this paper, our aim is to optimize the sensing performance\footnote{\textcolor{black}{Typically, the radar first estimates the angles of arrival of the targets by transmitting omnidirectional waveforms. The range is then estimated by transmitting waveforms towards these known target directions. In this paper, we focus on the latter case.}} given the constraints of communication qualities\footnote{The proposed scheme can be easily extended to the case where the communication performance is optimized, and thus we omit it in this paper.}. The sensing performance is promoted by minimizing the beampttern mismatch error, i.e., the difference between the transmit beampattern and a desired pattern, where the desired pattern has peaks in the target directions~(as in~\cite{stoica2007on}). Besides, we use channel capacity to evaluate the communication qualities between the RHS and the communication users.}

\textcolor{black}{It is challenging to solve this problem because the optimization of the digital and analog beamformers are coupled with each other, which is non-trivial. To efficiently handle this challenge, the optimization problem is first decoupled into two subproblems, i.e., the digital and the analog beamforming subproblems, where the digital/analog beamformer is optimized given the other. Next, a holographic beamforming optimization algorithm is designed to solve the optimization problem by iteratively optimizing the digital beamformer and the analog beamformer. Specifically, we first initialize the analog beamformer by summing the holographic patterns with equal weights. Next, the digital and the analog beamforming subproblems are sequentially solved in each iteration. The iteration terminates when the value difference of the beampattern mismatch errors~\cite{stoica2007on} between the two adjacent iterations is less than a predetermined positive constant $\epsilon$. In the following, we elaborate on the methods we use to tackle the two subproblems\footnote{\textcolor{black}{The running time of the proposed algorithm is less than $1$s in a moderate setting ($3$ users and $3$ targets), which is acceptable for scenarios with slow-moving targets or communication users such as pedestrians. Besides, when implementing the proposed ISAC system in practice, the computational tasks can be offloaded to the edge servers with greater processing power, which is able to reduce the computing time delay. In the future, a low-complexity and non-iterative algorithm can also be developed to support the real-time requirements in the fast-moving scenarios.}}}.

\textcolor{black}{\emph{Optimization of Digital Beamformer:} The digital beamformers for communication and radar sensing can be obtained by applying the zero forcing~(ZF) method~\cite{liu2020joint}. The basic idea is to first enforce the cancellation of the inter-user interference and the radar interference for all the communication users and then to derive the corresponding digital beamformers based on the channel information.}

\emph{Optimization of Analog Beamformer:} To optimize the weights in the analog beamformer, the subproblem is first transformed into a quadratic program by reformulating the objective function and the constraints in the subproblem. Next, the SDR technique can be applied to solve the quadratic problem.

\begin{figure}[!t]
  \setlength{\abovecaptionskip}{-0pt}
  \setlength{\belowcaptionskip}{-18pt}
  \centering
  \includegraphics[height=1.8in]{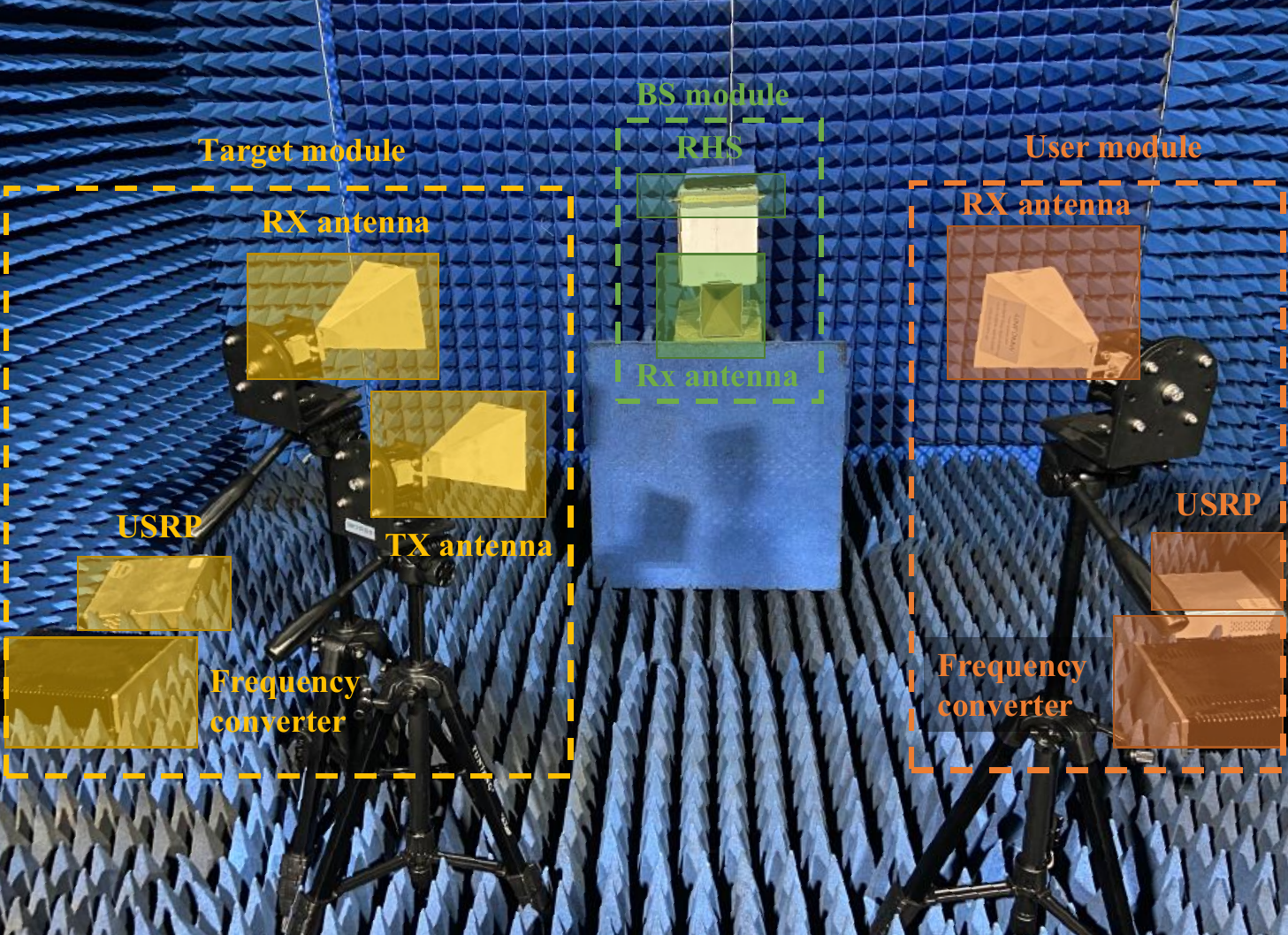}
  \caption{Experimental layout.}
  \label{f_experiment}
  \vspace{-2mm}
\end{figure}

\vspace{-2mm}
%%%%%%%%%%%%%%%%%%%%%%%%%%%%%%%%%%%%%%%%
\section{Prototype of the Holographic ISAC System}
\label{s_phis}
%%%%%%%%%%%%%%%%%%%%%%%%%%%%%%%%%%%%%%%%

In this section, we develop a hardware prototype of the holographic ISAC system. The implementation of the RHS is first described, and then the hardware modules which comprise the ISAC system are introduced.

\vspace{-2mm}
\subsection{Implementation of the RHS}
\label{ss_ir}

As shown in Fig.~\ref{f_prototype}, we design an 1D RHS whose dimension is $15 \times 3 \times 0.17$cm$^3$. The RHS consists of two SMA connectors, a multi-layer substrate, and 16 metamaterial elements. One of the SMA connectors serves as the feed which joins the RHS and the RF chain together. The other SMA connector joins the RHS and a $50\Omega$ RF load in order to absorb the energy remaining in the substrate. The substrate functions as a waveguide and is composed of four layers. The first and the third layers are made of F4B, and the second and the fourth layers are made of copper. The second layer of the substrate is the ground layer carrying a voltage of $0$V. The fourth layer is the DC feed line layer. There are $16$ feed lines, each connecting to a metamaterial element through a via hole. The other side of the feed line links to an output pin of the FPGA which applies a bias voltage to the element. 

The type of metamaterial elements arranged on the top of the substrate is complementary-electric-resonator (CELC), and two PIN diodes (MADP-000907-14020) are laid on each metamaterial element. Since the two PIN diodes are in parallel, they have the same voltage bias, which means each element can be tuned between two states, i.e., ON and OFF states. At the $11$GHz working frequency, the radiated energy of the element in the ON state is much greater than that in the OFF state, which forms the basis for the adjustment of the radiation pattern of the RHS.

\vspace{-2mm}
\subsection{Hardware Modules of the Holographic ISAC Prototype}
\label{ss_hmp}

The ISAC prototype is composed of three modules, i.e., the BS, user, and target modules, which are elaborated on below.

\subsubsection{ISAC transceiver module}

This module serves as an ISAC BS which transmits ISAC signals and receive echo signals for radar detection. To fulfill this task, an Intel NUC is implemented as the host computer. It controls the radiation amplitudes of the RHS elements via FPGA. It also connects with a USRP N210 which is able to simultaneously transmit and receive signals. Since the working frequency of the RHS ($12$GHz) is beyond the frequency range of the USRP ($0-6$GHz), a frequency converter is employed to up-convert the low-frequency signal transmitted by the USRP or down-convert the high-frequency signal received by the Rx antenna. The Rx antenna connecting to the frequency converter is a standard horn antenna (LB-75-20-C-SF) with a frequency range of $10-15$GHz.

\subsubsection{User module}

The user module receives and decodes the ISAC signal from the ISAC transceiver module to retrieve the communication stream. Specifically, the Rx antenna first receives the ISAC signal and sends it to the frequency converter. The frequency converter down-converts the received signal and transmits it to a USRP which down-converts the signal to the baseband. The baseband signal is finally sent to the NUC via Ethernet cable for decoding.

\subsubsection{Target module}

This module is used to simulate radar targets by generating controllable radar echo signals~\cite{ma2021spatial}. It consists of an RX antenna, a TX antenna, a frequency converter, a USRP, and an Intel NUC. Once the Rx antenna receives the ISAC signal transmitted by the RHS, the target module is triggered, which adds delays to the ISAC signal and emits the delayed signal through the TX antenna. The value of the delayed time can be adjusted by the PC application running on the Intel NUC in order to simulate the targets located at different ranges.

%%%%%%%%%%%%%%%%%%%%%%%%%%%%%%%%%%%%%%%%
\section{Performance Evaluation}
\label{s_pe}
%%%%%%%%%%%%%%%%%%%%%%%%%%%%%%%%%%%%%%%%

\begin{figure}[!t]
  \setlength{\abovecaptionskip}{-0pt}
  \setlength{\belowcaptionskip}{-18pt}
  \centering
  \includegraphics[height=1.2in]{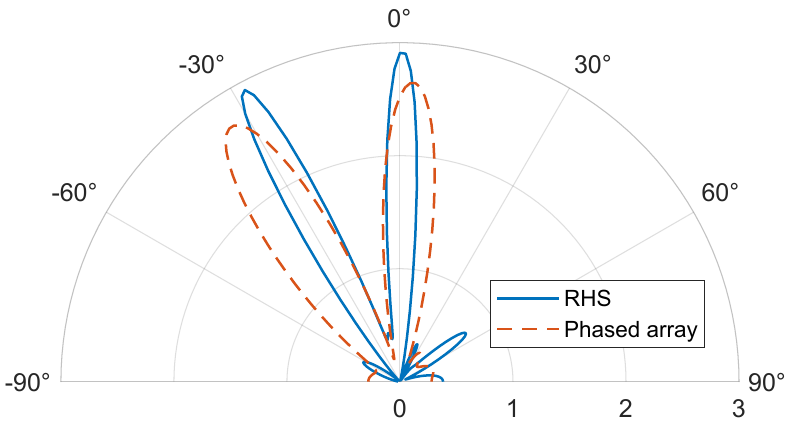}
  \caption{Radiation patterns of the RHS and phased array.}
  \label{f_pattern}
  \vspace{-2mm}
\end{figure}

\begin{figure*}[!t]
  \setlength{\abovecaptionskip}{-0pt}
  \setlength{\belowcaptionskip}{-18pt}
  \centering
    \label{a1}
  \includegraphics[height=2in]{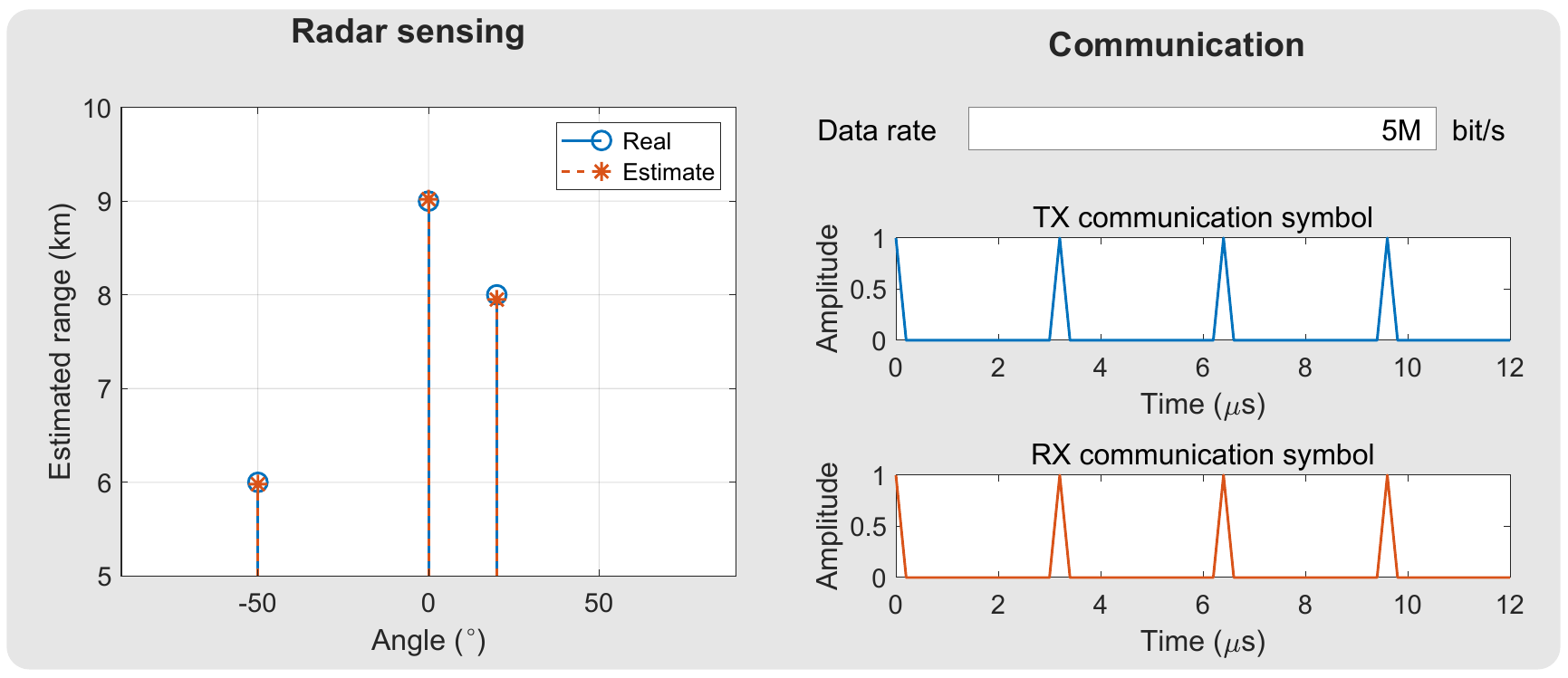}
  \caption{Radar sensing and communication performances of the holographic ISAC system.}
  \label{f_result}
  \vspace{-2mm}
\end{figure*}

In this section, we evaluate the performance of the proposed ISAC system. The experiment layout is shown in Fig.~\ref{f_experiment}. We deploy the proposed prototype in an anechoic chamber with a size of $4 \times 3 \times 2.5$m$^3$. The ISAC transceiver module is located at the center of the chamber. The user and the target modules are placed in different directions in regard to the ISAC transceiver module, and the distance between the ISAC transceiver module and the user/target module is $1.7$m. In the ISAC transceiver module, the ISAC signal transmission and the radar signal reception are performed in a time-division manner. Specifically, the BS first transmits the ISAC signal with a duration of $12\mu$s and then listens for the echo signal to decide the presence of the target. Since there is only one RF chain in the BS, the digital beamformer is fixed as 1, and the analog beamformer is optimized to promote the ISAC performance.

Fig.~\ref{f_pattern} shows the radiation patterns of the RHS and the phased array. The phased array contains $5$ antenna elements, whose relative phases can be independently tuned by phase shifters. To simultaneously detect the target and serve the communication user, the RHS or the phased array generates two beams that point towards directions $0^\circ$ and $-30^\circ$. We can observe that the gains of the RHS towards the directions of interest are slightly higher than those of the phased array. Besides, the power consumption of the phased array ($5$W) is substantially larger than that of the RHS ($0.16$W), which indicates that the RHS is able to support ISAC with similar performance and lower consumption compared with the phased array.

Fig.~\ref{f_result} illustrates the ISAC performances of the proposed platform. We place the target module in directions $-50^\circ$, $0^\circ$, and $20^\circ$, respectively, to simulate the targets in different directions. The delays added to the ISAC signal by the target module are set as $20\mu$s, $30\mu$s, and $26.7\mu$s which corresponds to the target in $6$km, $9$km, and $8$km away. In order to sense the target, one of the main lobes of the radiation pattern is steered towards directions $-50^\circ$, $0^\circ$, and $20^\circ$ in different cycles, while the other main lobe keeps pointing toward the direction of the user module, i.e., $60^\circ$, to support downlink communication. It can be observed from Fig.~\ref{f_result} that the estimated range is close to the real range, which proves the feasibility of sensing by applying holographic ISAC. Besides, we can also observe from Fig.~\ref{f_result} that the communication symbol received by the user module is the same as the communication symbol transmitted by the BS, and the data rate between the BS and the user is $5$M bit/s, which shows that the communication between the BS and the user can be supported when the BS performs radar sensing at the same time.

%%%%%%%%%%%%%%%%%%%%%%%%%%%%%%%%%%%%%%%%
\section{Future Research Directions and Key Challenges}
\label{s_frdkc}
%%%%%%%%%%%%%%%%%%%%%%%%%%%%%%%%%%%%%%%%

In the previous sections, we have introduced the concept of holographic ISAC and shown the potential benefit of this concept compared to a traditional phased array. In this section, we introduce future research directions for RHS-based ISAC and the corresponding key challenges.

\subsection{Fundamental Designs of the RHS}

The designs of RHS layout parameters are critical in the optimization of holographic ISAC systems. In the following, we introduce the designs of two vital parameters, i.e., the RHS size and element spacing.

\begin{itemize}
  \item \textbf{Design of RHS size:} To meet the increasing demand for communication capacity and sensing accuracy, the size of the RHS needs to be enlarged to provide higher antenna gain. However, the signal attenuation in the waveguide cannot be ignored for large-scale RHSs, which significantly decreases the efficiency of the antenna. Consequently, the size of the RHS should be optimized to balance radiation efficiency and antenna gain.
  \item \textbf{Design of element spacing:} Since a narrow spacing between two nearby RHS elements leads to a large number of elements given the size of the antenna aperture, the directionality of the RHS can be improved by reducing the element spacing. However, the element spacing cannot be unlimitedly decreased because the effect of mutual coupling increases when the element spacing decreases, which degrades the radiation performance of the RHS.
  \item \textcolor{black}{\textbf{Scaling to higher frequencies:} Driven by higher data rate requirements and the accommodation of more users, the communication systems are moving towards an unused spectrum with higher frequencies and larger bandwidth such as millimeter wave~(mmWave). The proposed ISAC scenario also has the potential for the scaling to higher frequencies, where the RHS can be easily integrated with the mmWave circuits to reduce the profile and weight of the system. Novel designs of metamaterial structure are required to enable a high antenna gain of the RHS in the mmWave band to compensate the severe attenuation of mmWave transmission.}
\end{itemize}

\subsection{Limitations and trade-offs of Holographic ISAC}

It is essential to theoretically analyze the limitations and trade-offs to further verify the superiority of holographic ISAC. To provide a general framework for the analysis, a performance bound which unifies both radar and communication is necessary. 

For traditional ISAC systems, many existing works are devoted to developing closed-form expressions of the performance bounds by exploiting the inherent relation between information theory and detection theory~\cite{liu2020jointradar}. However, due to the differences in hardware structures and working principles, the performance bounds developed for traditional ISAC systems cannot be directly applied to holographic ISAC. A new performance bound needs to be developed for RHS-based ISAC systems.

\subsection{Optimization of Holographic ISAC Transceiver}

As a metamaterial antenna, the RHS can also be utilized for signal reception, indicating that the concept of holographic ISAC can be extended to the scenario where both the TX and RX antennas of the BS are replaced by the RHSs. However, several challenges lie in designing such a scheme. 
\begin{itemize}
  \item \textcolor{black}{\textbf{Channel Estimation:} Channel information is critical to the optimization of communication performances, while it is challenging to estimate the communicaiton channels due to the hybrid analog-digital strucutre of the RHSs. A straightforward method is to leverage the amplitude-controllable capability of the RHSs and estimate the channel of each element in a time division manner. Specifically, in each time slot, only one element is turned on, i.e., the radiation amplitude of this element is maximized. The radiation amplitudes of other elements are set as $0$ in order to avoid the interference caused by these elements. The time overhead of such method is proportional to the size of the RHS, indicating the time cost may become relatively large for large-scale RHSs. Thus, novel method that can simultaneously estimate the channels of multiple elements should be developed to reduce the corresponding burden of channel estimation.}
  \item \textcolor{black}{\textbf{AoA Estimation:} Considering the unique structure of the RHS where signals are first modulated by the superposition of holographic patterns and then received by the feeds, the estimation of angles of arrivals~(AoAs) by using RHSs is more complicated than that by using fully-digital arrays where the signals are directly received by the feeds. To address this issue, the signals that arrive at the RHS from unknown AoAs should be modeled, and the maximum likelihood principle can be applied to estimate the AoAs based on the model and the signals received by the RHS feeds. To reduce the related computational overhead, an efficiently estimation algorithm is required to perform multi-dimensional search and find the optimal solution of AoAs.}
  \item \textbf{Joint Design:} In the design of a holographic ISAC transceiver, the beamformers for transmission and the combiners for reception need to be jointly optimized, which is much more complicated than solely optimizing the beamformers or the combiners.
\end{itemize}

\section{Conclusion}
\label{s_c}
In this article, we have introduced RHS-enabled holographic ISAC, a new paradigm for integrating sensing and communication functions. We have presented the concept of RHSs and developed a hybrid beamforming scheme for holographic ISAC based on the working principle of RHSs. In particular, we have built a hardware prototype of the RHS-enabled holographic ISAC system. Simulation and experimental results have shown that the power consumption of the RHS is lower than that of the phased array with a similar antenna gain, and unveiled the great potential for energy saving by implementing holographic ISAC. We have also discussed future research direction and key challenges of holographic ISAC.

\bibliographystyle{IEEEtran}
\bibliography{IEEEabrv,myReference}

\newcommand{\BioSpace}{-12}

\vspace{10mm}

\section*{Biographies}

\vspace{\BioSpace mm}

\begin{IEEEbiographynophoto}{Haobo Zhang} [S’19] (haobo.zhang@pku.edu.cn) received his B.S. degree in electronic engineering from Peking University, China, in 2019, where he is currently pursuing a Ph.D. degree with the Department of Electronics.
\end{IEEEbiographynophoto}

\vspace{\BioSpace mm}

\begin{IEEEbiographynophoto}{Hongliang Zhang} [M’19] (hongliang.zhang92@gmail.com) is a postdoctoral associate in the Department of Electrical and Computer Engineering at Princeton University. He was the recipient of the 2021 IEEE Comsoc Heinrich Hertz Award.
\end{IEEEbiographynophoto}

\vspace{\BioSpace mm}

\begin{IEEEbiographynophoto}{Boya Di} [M’19] (diboya92@gmail.com) is an assistant professor at the Department of Electronics, Peking University. She serves as an Associate Editor of IEEE Trans. Vehicular Technology.
\end{IEEEbiographynophoto}

\vspace{\BioSpace mm}

\begin{IEEEbiographynophoto}{Lingyang Song} [F’19] (lingyang.song@pku.edu.cn) joined the Department of Electronics, Peking University in May 2009, where he is currently a Boya Distinguished Professor. He was a recipient of the IEEE Leonard G. Abraham Prize in 2016 and the IEEE Asia Pacific (AP) Young Researcher Award in 2012. He has been an IEEE Distinguished Lecturer since 2015.
\end{IEEEbiographynophoto}

\end{document}